\documentclass[12pt,preprint,useAMS,usenatbib]{emulateapj}
\usepackage{graphicx}
\usepackage{amssymb}
\usepackage{amsmath}
\usepackage{threeparttable}

\def\msun{\,{\rm M_\odot}}

\def\spose#1{\hbox to 0pt{#1\hss}}
\def\lta{\mathrel{\spose{\lower 3pt\hbox{$\mathchar"218$}}
     \raise 2.0pt\hbox{$\mathchar"13C$}}}
\def\gta{\mathrel{\spose{\lower 3pt\hbox{$\mathchar"218$}}
     \raise 2.0pt\hbox{$\mathchar"13E$}}}

\newcommand{\etal}{{et al.\ }}
\newcommand{\nar}{{NewAR}}

\def\kms{\,{\rm km\,s^{-1}}}

\lefthead{Rashkov \& Madau}
\righthead{Relic IMBHs in the halo of the Milky Way}
\submitted{submitted to the ApJ}


\begin{document}

\title{A population of relic intermediate-mass black holes in the halo of the Milky Way}
\author{Valery Rashkov and Piero Madau}
\affil{Department of Astronomy and Astrophysics, University of California, Santa Cruz, 1156 High Street, Santa Cruz, CA 95064.}

\begin{abstract} 
If ``seed" central black holes were common in the subgalactic building blocks that merged to form present-day massive galaxies, 
then relic intermediate-mass black holes (IMBHs) should be present in the Galactic bulge and halo. We use a particle tagging 
technique to dynamically populate the N-body {\it Via Lactea II} high-resolution simulation with black holes, and assess the size, 
properties, and detectability of the leftover population. The method assigns a black hole to the most tightly bound central 
particle of each subhalo at infall according to an extrapolation of the $M_{\rm BH}-\sigma_*$ relation, and self-consistently 
follows the accretion and disruption of Milky Way progenitor dwarfs and their holes in a cosmological ``live" host from high 
redshift to today. We show that, depending on the minimum stellar velocity dispersion, $\sigma_m$, below which central black holes 
are assumed to be increasingly rare, as many as $\sim 2000$ ($\sigma_m=3\,\kms$) or as few as $\sim 70$ ($\sigma_m=12\,\kms$) IMBHs may be 
left wandering in the halo of the Milky Way today. The fraction of IMBHs kicked out of their host by gravitational recoil is $\lta 20$\%.
We identify two main Galactic subpopulations, ``naked" IMBHs, whose host subhalos were totally destroyed after infall, and ``clothed" IMBHs 
residing in dark matter satellites that survived tidal stripping. Naked IMBHs typically constitute 40-50\% of the total and are more centrally 
concentrated. We show that, in the $\sigma_m=12\,\kms$ scenario, the clusters of tightly bound stars that should accompany naked IMBHs would 
be fainter than $m_V=16$ mag, spatially resolvable, and have proper motions of 0.1--10 milliarcsec per year. Their detection may provide an 
observational tool to constrain the formation history of massive black holes in the early Universe. 
\end{abstract}
\keywords{black hole physics -- Galaxy: halo -- stellar content -- galaxies: evolution -- dwarf -- method: numerical}

\section{Introduction} \label{intro}

Direct dynamical measurements show that most local massive galaxies host a quiescent massive black hole (MBH) in their nuclei. 
Their masses have been found to correlate tightly with the mass \citep{har04} and the stellar velocity dispersion of the host stellar bulge, as manifested in the $M_{\rm BH}-\sigma_*$ relation of spheroids \citep{fer00, geb00}. 
It is not yet understood whether such scaling relations were set in primordial structures and maintained throughout cosmic time with a small dispersion, or indeed which physical processes established such correlations in the first place. It is also unclear whether there exists a minimum host galaxy mass or velocity dispersion below which MBHs are unable to form or grow. 
The subset of the MBHs that populates the centers of dwarf galaxies and very late spirals and is undergoing accretion can be detected as active galactic nuclei (AGNs). Observations of AGNs with estimated black hole masses in the range $10^5$-$10^6\,\msun$ appear to be consistent with an extrapolation of the local $M_{\rm BH}-\sigma_*$ relation for inactive galaxies down to stellar velocity dispersions of $30\,\kms$ \citep{bar05, xia11}.
The existence of ``intermediate-mass" black holes (IMBHs) with $20\,\msun<M_{\rm BH}<10^5\,\msun$ remains in dispute. Dynamical mass measurements have shown the presence of IMBH candidates in the cores of the globular clusters G1 \citep[$M_{\rm BH}=1.8\pm 0.5\times 10^4\,\msun$,][]{geb05}, $\omega$ Centauri \citep[$M_{\rm BH}=4.7\pm 1.0\times 10^4\,\msun$,][]{noy10}, NGC 1904 and 6266 \citep[$M_{\rm BH}=3\pm 1\times 10^3\,\msun$ and $M_{\rm BH}=2\pm 1\times 10^3\,\msun$,][]{lut12}.
The X-ray spectra and bolometric luminosities of the ultraluminous off-nuclear X-ray sources detected in nearby galaxies may imply the presence of IMBHs with $M_{\rm BH}\gta 500\,\msun$ \citep{far09, kaa01}.    

While the ``seeds" of the MBHs powering the $z\gta 6$ {\it Sloan Digital Sky Survey} (SDSS) very luminous, rare quasars \citep{fan06} must have appeared at very high redshifts and grown rapidly to more than $10^9\,\msun$ in less than a Gyr, the problem of their 
origin and occupation fraction in early galaxies remains unsolved. MBHs may have grown from the remnants of Population III (Pop III) star 
formation in sub-dwarf galaxies at $z\gta 15$ \citep[e.g.][]{mad01,vol03,tan09}, or from more massive precursors formed by the ``direct collapse" of large amounts of gas in dwarf galaxy systems at later times \citep{loe94, kou04, beg06, lod06, may10}. Massive seeds have larger masses than Pop III remnants, but form in rarer hosts. Questions remain in the Pop III remnant model about the ability of $\sim 100\,\msun$ seed holes to grow at the Eddington rate for tens of e-folding times unimpeded by feedback, and in the direct collapse model about the needed large supply of low angular momentum gas that must accumulate in the center of young galaxies before fragmentation and star formation sets in (see, e.g., Haiman 2012 and references therein).

In this Paper we assess the size, properties, and detectability of the leftover population of IMBHs\footnote{In the following, we will use the 
term IMBH to refer to any hole with $20\,\msun<M_{\rm BH}<10^5\,\msun$, i.e. more massive than the stellar-mass black holes found 
in X-ray emitting binary systems \citep{oro07} and 40 times less massive than Sgr A$^*$.}\, that is predicted to survive today in the 
halo of the Milky Way (MW) galaxy by two fiducial seeding scenarios. It was first pointed out by \citet{mad01} that, if seed holes were 
indeed common in the subgalactic building blocks that merged to form the present-day massive galaxies, then a numerous population of 
relic IMBHs should be present in the Galactic bulge and halo \citep[see also][]{vol03, isl03, vol05}. \citet{mic06, mic11} used 
collisionless simulations to study the effect of gravitational recoil kicks on the IMBH distribution in present-day galaxies and test 
merger-driven recipes for black hole growth. \citet{van10} ran Monte Carlo realizations of the merger history of massive galaxy halos to study 
IMBHs in MW satellites at $z=0$. A similar approach was employed by \citet{ole09} to estimate the expected number of recoiled IMBH remnants 
present today in the MW halo. Cosmological hydrodynamic simulations of the growth of MBHs are either limited by resolution to galaxy hosts 
with $M_{\rm halo}\gta 10^{10}\,\msun$ \citep[e.g.][]{dim08,dub12} or stop at high redshift \citep[e.g.][]{bel11}. In all approaches seed holes 
are planted following prescriptions that are based either on local properties such as gas angular momentum, temperature, metallicity, etc, or on 
global properties such as the mass or circular velocity of the host halo. To complement the above calculations, we use here a {\it particle tagging 
technique} to dynamically populate the N-body {\it Via Lactea II} (VLII) extreme-resolution simulation with IMBHs. As we shall discuss, this method 
allows us to self-consistently follow the accretion and disruption of thousands of MW progenitor dwarfs and the kinematics of their holes in 
a cosmological ``live" host, and the build-up of a large population of Galactic ``naked" wandering IMBHs.

\section{Black Hole Tagging Technique} \label{sec:simmodel}

The cosmological $\Lambda$CDM VLII simulation, one of the highest-precision N-body calculations of the assembly of the Galactic halo to date \citep{die08}, was performed with the PKDGRAV tree-code \citep{sta01}. It employs just over one billion $4,100\,\msun$ particles to model the formation of a $M_{200}=1.93\times 10^{12}\,\msun$ Milky Way-sized halo and its substructure. About 20,000 surviving subhalos of masses above $10^6\,\msun$ are resolved today within the main host's $r_{200}=402$ kpc (the radius enclosing an average density 200 times the mean matter value). Central black holes are added to subhalos following the particle tagging technique detailed in \citet{ras12} and quickly summarized here. In each of 27 (out of the 400 available) snapshots of the simulation, chosen to span the assembly history of the host between redshift $z=27.54$ and the present, all subhalos are identified and linked from snapshot to snapshot to their most massive progenitor: the subhalo tracks built in this way contain all the time-dependent structural information necessary for our study. We then: 1) identify the simulation snapshot in which each subhalo reaches its maximum mass, $M_{\rm halo}$, before being accreted by the main host and tidally stripped; 2) measure the subhalo maximum circular velocity $V_{\rm max}$; 3) link it to the stellar line-of-sight velocity dispersion, $\sigma_*$, using the relation $V_{\rm max}=2.2\sigma_*$ derived by \citet{ras12}; and 4) tag {\it the most tightly bound central particle} as a black hole of mass $M_{\rm BH}$ according to an extrapolation of the $M_{\rm BH}-\sigma_*$ relation of \citet{tre02},  
\begin{equation}
{M_{\rm BH}\over \msun}= 
\begin{cases} 10^{6.91} \left(\frac{\sigma_*}{100\,\kms}\right)^4  & (\sigma_*\ge 6\,\kms) 
\\ 
100 & (\sigma_*< 6\,\kms). \end{cases}
\label{msigma}
\end{equation}
A significantly steeper $M_{\rm BH}-\sigma_*$ relations have been derived recently by, e.g. \citet{gra11} and \citet{mcc13}, and we will discuss 
the impact of such a steeper power-law on our results in \S\,6. 
By neglecting all the poorly resolved subhalos with $M_{\rm halo}<10^7\,\msun$, we restrict our analysis to 3,204 such tracks. 
{Note that, with such a cut, about 200 subhalos below $10^7\,\msun$ and with $\sigma_*>3\,\kms$ (equivalent to $V_c > 6.6\,\kms$) are not actually assigned
a central black hole. These ``missed" IMBHs account, however, for less than 10\% of all possible IMBHs that are tagged at these low stellar 
velocity dispersions.}

Any evolution of the tagged holes after infall is purely kinematical in character, as their satellite hosts are accreted and disrupted in an evolving Milky Way-sized halo. After tagging, the black hole particles are tracked down to the $z=0$ snapshot. The main host is assigned a central black hole at $z=0$ of mass equal to that of Sgr A$^*$, $M_{\rm BH}=4\times 10^6\,\msun$ \citep{ghe08}. While we tag at most one hole per subhalo, we allow multiple systems to form when subhalos merge after infall (see below).
Figure \ref{fig1} shows the distributions of stellar velocity dispersion at infall (left panel) and time since infall (right panel) for the VLII
subhalo population. The color coding separates self-gravitating subhalos that survive their accretion event and become MW satellites 
from those that are totally disrupted. The latter are found to fall in preferentially at earlier times, and have a median infall 
redshift of $4.5$. 

\begin{figure*}
\centering
\includegraphics[width=0.48\textwidth]{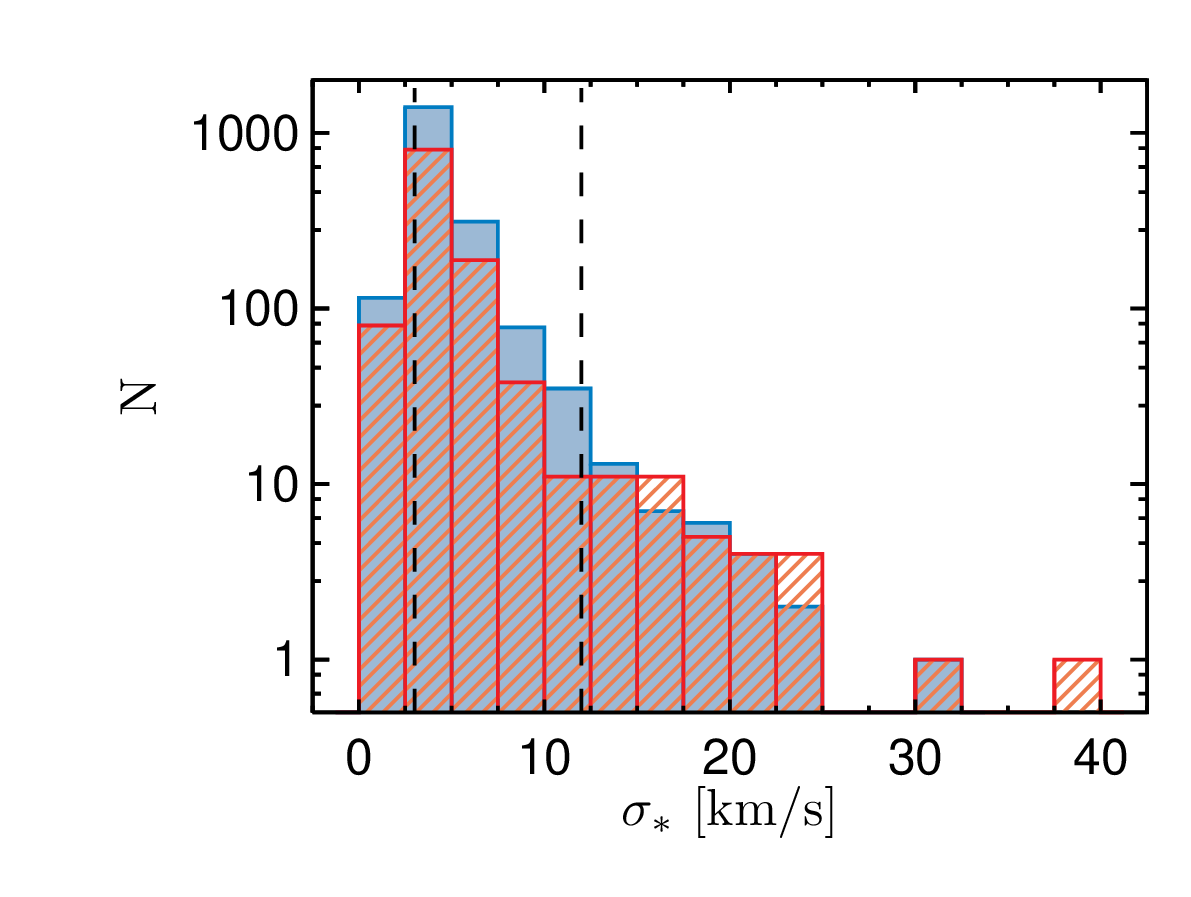}
\includegraphics[width=0.48\textwidth]{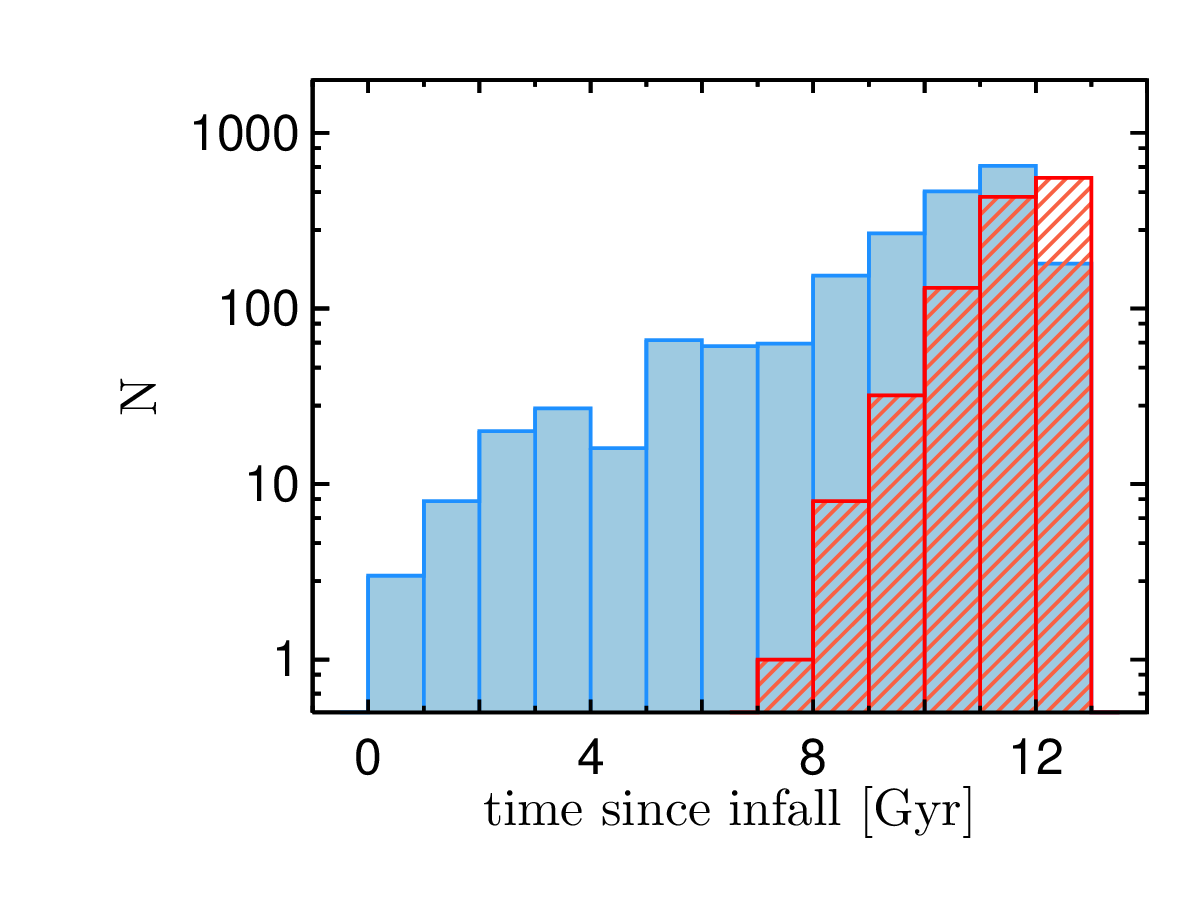}
\caption{Histograms of the stellar velocity dispersion at infall ({\it left panel}) and infall lookback time  
({\it right panel}) for all subhalos with $M_{\rm halo}>10^7\,\msun$ in the VLII simulation. The color coding 
separates self-gravitating subhalos that survive their accretion event and become MW satellites ({\it blue}) 
from those that are totally disrupted ({\it red}). The vertical lines in the left panel mark the minimum velocity dispersion
below which central black holes are assumed to be increasingly rare in the two seeding scenarios discussed in the text.
}
\label{fig1}
\vspace{+0.3cm}
\end{figure*}

\section{Demography of IMBHs}

Below we discuss two simple models of seed hole formation that may be illustrative of more realistic growth scenarios, and differ only in the black hole occupation fraction as a function of the host's stellar velocity dispersion.  

\subsection{Population III Remnants}

Let us first consider a scenario where the black hole occupation fraction is of order unity at infall in all subhalos with stellar velocity 
dispersions $\ge \sigma_m=3\,\kms$ and drops to zero below $\sigma_m$. This value of $\sigma_m$ is comparable to the stellar velocity 
dispersion measured today in the ultra-faint MW satellite Segue 1 \citep{sim11}, i.e. this model places seed holes in small-mass subhalos that are 
known to have been rather inefficient at forming stars \citep[e.g.][]{ras12,coo10,kop09}. {The mass distribution of subhalos with 
$\sigma_*>3\,\kms$ at infall has a median of $5\times 10^7\,\msun$, so these subhalos are well resolved in our simulation. We assign black holes masses 
following the relation (\ref{msigma}). As most of the subhalos in Fig. \ref{fig1} host then a $100\,\msun$ black hole, we refer to this 
model as ``Pop III remnants".  We note that this scenario is not based on a specific physical model, but is simply meant to be used for comparison 
with existing and forthcoming numerical simulations. The occupation fraction is arbitrarily set to unity but can be easily scaled down to account
for a reduced MBH formation efficiency on these small subhalo mass scales.}

Two large subpopulations of Galactic black holes can then be readily identified today within $r_{200}$, 798 ``naked" IMBHs, whose host subhalos 
have been totally disrupted after infall, and 1,234 ``clothed" IMBHs residing in satellites that survived tidal stripping. The abundance of Galactic 
black holes in this model is comparable with the predictions of semi-analytical calculations by \citet{isl04} for Milky Way-sized halos 
(their Model A). These authors also identified a population of ``naked" IMBHs following the tidal disruption of infalling satellites, but 
their estimated naked fraction appears smaller than we find in our simulations (see Fig. 2 of \citealt{isl04}). 
As expected, in both calculations (as well as in those of \citealt{vol05}) there are more naked holes near the center of the host, where tidal forces are stronger. 
Indeed, within 10 kpc of the center, most IMBHs are naked.

\begin{figure*}
\begin{center}
$\begin{array}{c@{\hspace{1mm}}c@{\hspace{1mm}}c}
\includegraphics[width=0.45\textwidth]{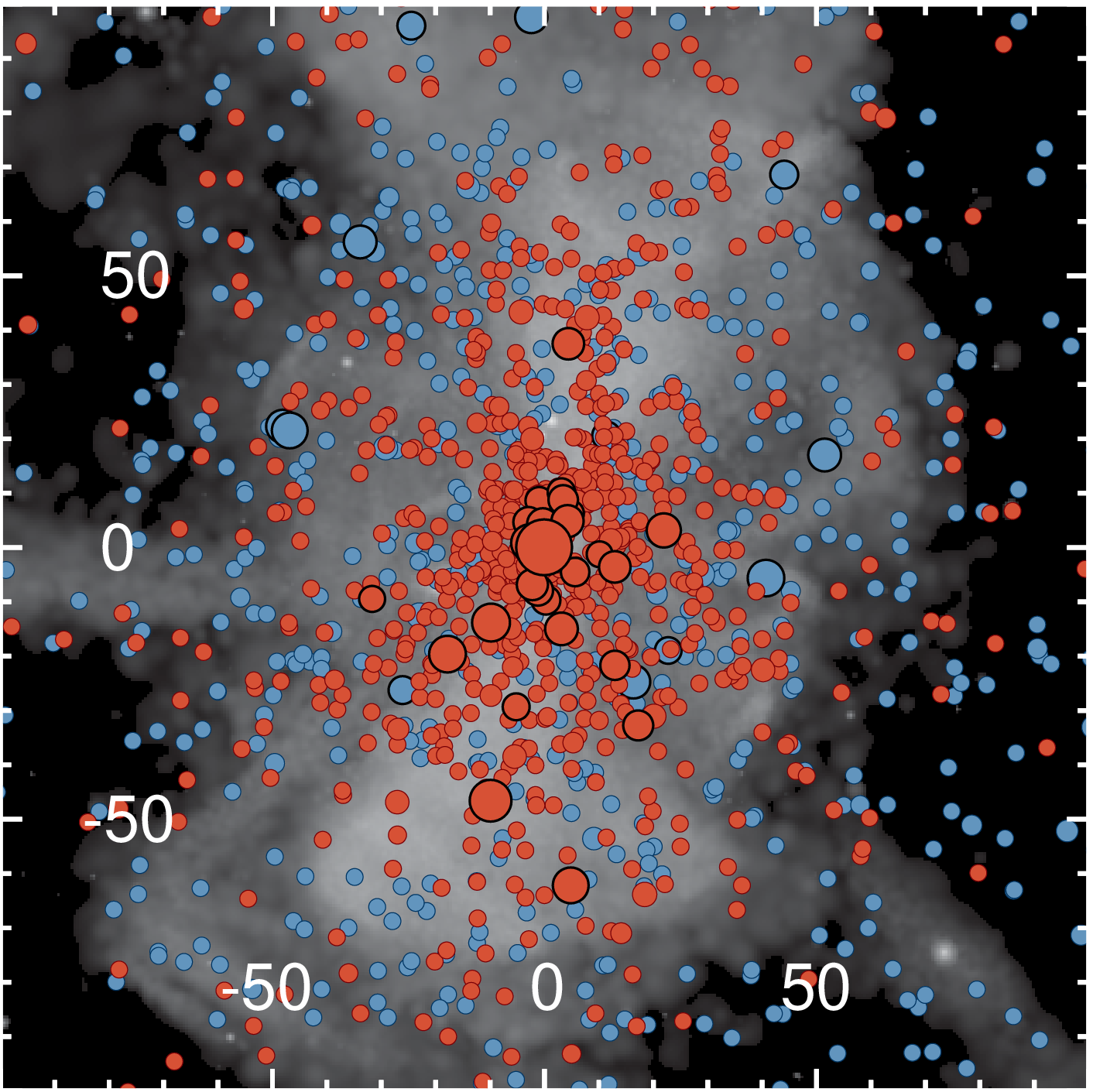}
\includegraphics[width=0.45\textwidth]{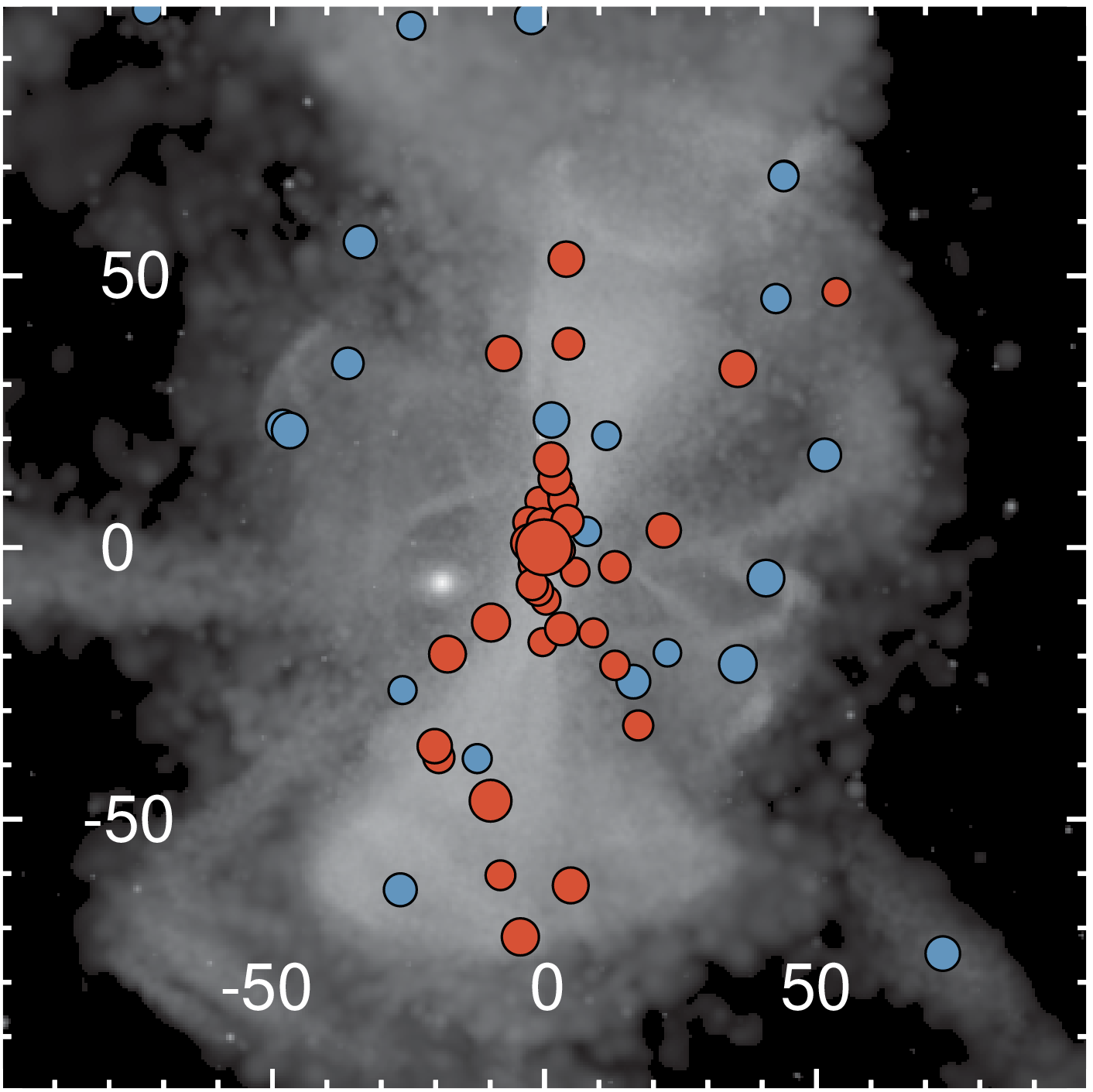}\\ 
\includegraphics[width=0.45\textwidth]{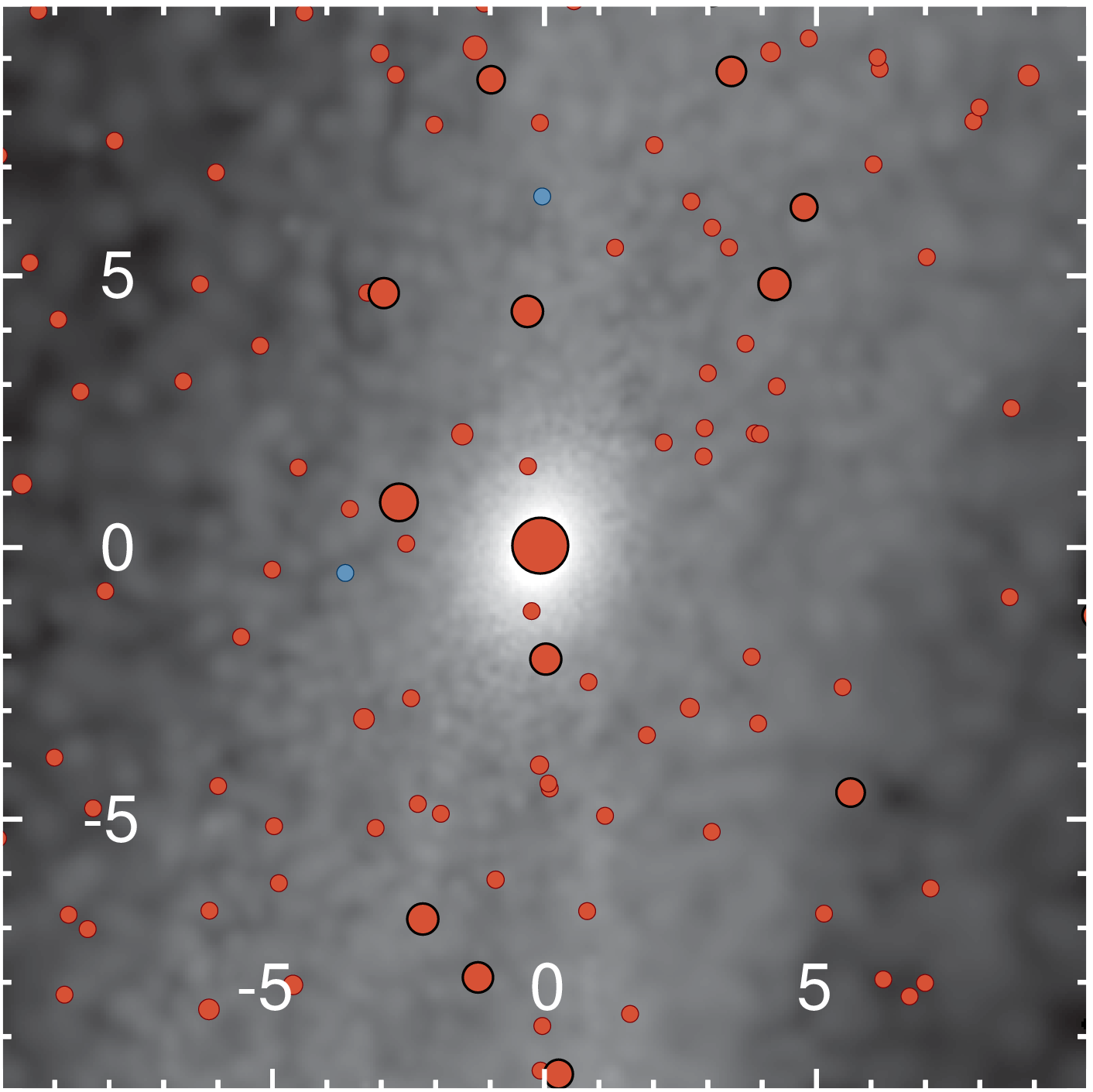} 
\includegraphics[width=0.45\textwidth]{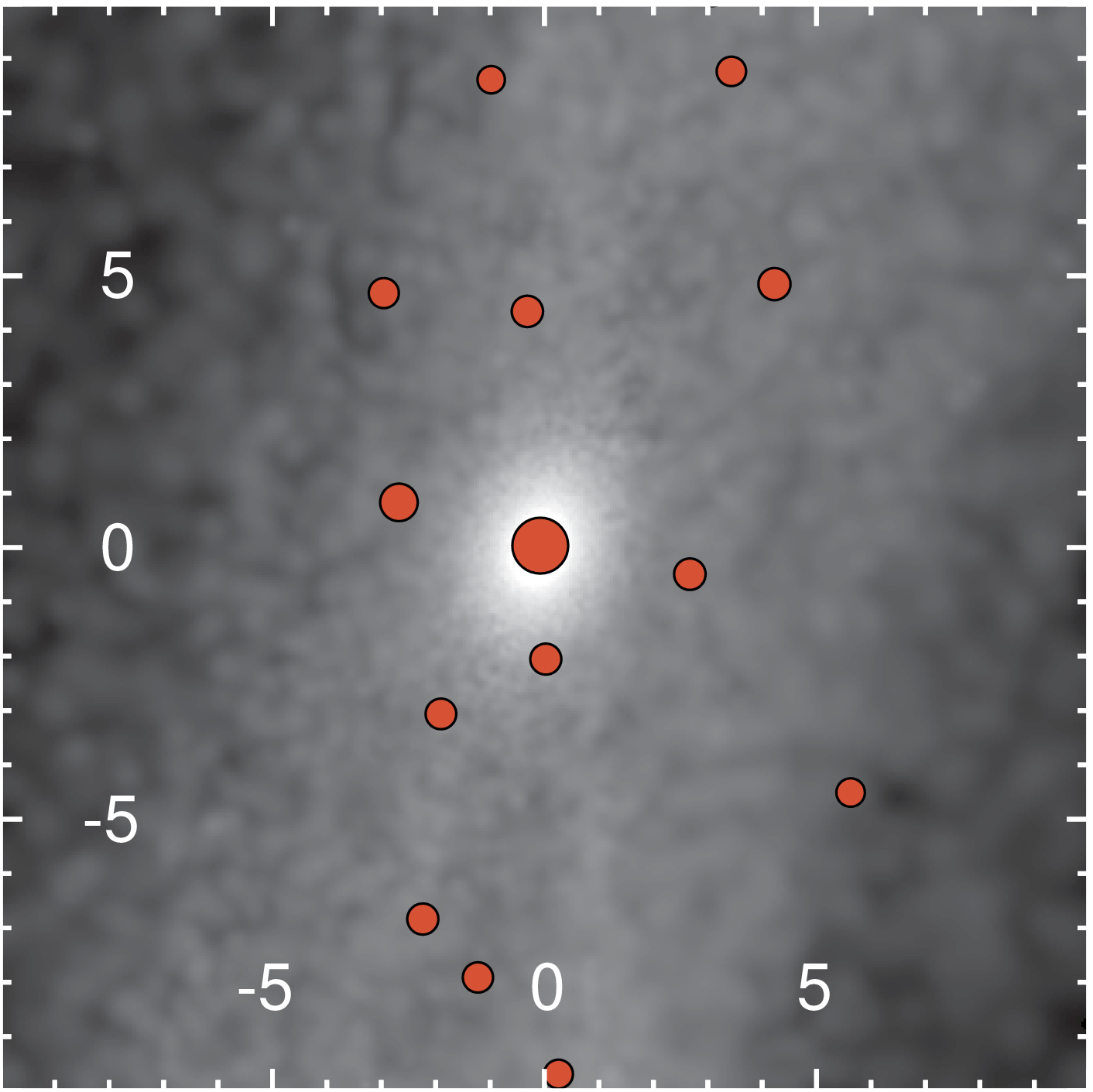} 
\end{array}$
\end{center}
\caption{Projected distribution of IMBHs in the Galactic halo today. {\it Left panels:} Pop III remnant model.  {\it Right panels:} Direct 
collapse progenitor model. The color coding is as in Fig. \ref{fig1}, while the size of the points is proportional to $\log_{\rm 10} M_{\rm BH}$. 
The stellar halo from \citet{ras12} is plotted in the background. {\it Top row:} 200 kpc box. {\it Bottom row:} Zoom-in of the inner 20 kpc region. 
}
\label{fig2}
\vspace{+0.3cm}
\end{figure*}

Within the ``clothed" category, 1,096 are single black hole systems, i.e. are located in subhalos that host 
only one hole at the present epoch. The others are multiple systems: we find 42 dwarf subhalos with 2 holes, 7 with 3, 5 with 4, and 1 with each of 6 and 7 holes. These multiple systems 
are produced when their host subhalos merge together after infall, as they fall into the main halo in groups. We assume that by that time, ram pressure stripping has removed any leftover 
gas from the system, making it difficult for IMBHs to sink to the center of the potential well and merge. Substructures that host 5 or more IMBHs are found to be more than 250 kpc 
from the Galactic center and have masses in the range $2\times 10^9-3\times 10^9\,\msun$ today. Figure \ref{fig2} (left panels) shows an image of the projected spatial distribution of IMBHs 
in today's inner and outer Galactic halo according to this model, while Figure \ref{fig3} depicts their mass function and radial distribution. Naked holes are more concentrated towards the 
inner halo regions as a consequence of the tidal disruption of infalling satellites. Indeed, within 10 kpc of the center, most IMBHs are naked. The second most massive hole 
(after ``Sgr A$^*$'') has $M_{\rm BH}=8.6\times 10^4\,\msun$ (i.e. was tagged at infall in a large satellite with $\sigma_*=32\,\kms$), is naked, and is located at a distance of 55 kpc from 
the center. There are 48 IMBHs above $M_{\rm BH}=10^3\,\msun$, of which 29 are naked. Most holes are assigned the seed mass of $100\,\msun$, i.e. they are not allowed to grow in this model: 
we count a total of 1,801 such ``light" IMBHs, of which 695 are naked.

\subsection{Direct Collapse Precursors}

Consider by contrast a scenario where more massive seeds form in rarer hosts. Models in which seed black holes become very rare
for halos with $M_{\rm halo}\lta 10^9\,\msun$ have been recently discussed by, e.g., \citet{bel11} and \citet{dev12}.
Here, we take the hole occupation fraction to be of order unity in all subhalos with stellar velocity dispersions $\ge \sigma_m=12\,\kms$, 
and to drop to zero below $\sigma_m$. Black hole masses are again assigned following equation (\ref{msigma}) down to $\sigma_m$. The mass 
distribution of subahalos hosting central black holes at infall now has a median of $3\times 10^9\,\msun$, and the minimum hole mass is 
$M_{\rm BH}\simeq 1,700\,\msun$. {Again, this scenario is not based on a specific physical model but is simply meant to be illustrative
of a situation in which dwarf spheroidal satellites of the Milky Way with stellar velocity dispersions comparable to that of Ursa Minor 
(see, e.g., \citealt{wol10} for a compilation) would actually host an IMBH. Recent observations have shown that dwarf galaxies with stellar masses 
comparable to the Magellanic Clouds do exhibit optical spectroscopic signatures of accreting MBHs \citep{rei13}. Note that, in the models 
of \citet{bel11}, halos with virial masses above $3\times 10^9\,\msun$ are predicted to always host a MBH seed, regardless of 
the seed formation efficiency. This is because, even in the lowest efficiency case, MBH seeds form in the region of earliest star formation, which
tend to be the halos that becomes the most massive later on in every simulation. Moreover, the most massive halos have also experienced the 
greatest number of mergers, and are then populated by MBH seeds brought in by satellites.}    

The tagging prescription adopted in this scenario gives origin to 39 naked and 33 clothed IMBHs (of which 27 are single systems). 
Figure \ref{fig2} (right panels) shows an image of the projected spatial distribution of IMBHs in today's inner and outer Galactic halo according to this model, 
while Figure \ref{fig3} (left panel) depicts their mass function. There are 20 IMBHs above $M_{\rm BH}=10^4\,\msun$, of which 13 are naked. 

\section{Black Hole Recoils}

The asymmetric emission of gravitational waves produced during the coalescence of a binary black hole system imparts a velocity kick to the system that can displace the hole from the center of its host. The magnitude of the recoil will depend on the binary mass ratio and the direction and magnitude of their spins \citep{bak08}, but not on the total mass of the binary. When the kick velocity is larger than the escape speed of the host halo, a hole may be ejected into intergalactic space before becoming a Galactic IMBH. The demography of IMBHs discussed in the previous section accounts for recoiling holes as follows.   

Following the results of high-resolution self-consistent gasdynamical simulations of binary mergers of disk galaxies by \citet{kaz05}, we assume that unequal-mass galaxy mergers with mass ratios larger than 4:1 will not lead to the formation of a close black hole pair at the center of the remnant, and hence of a recoiling hole. All major mergers (mass ratios smaller than 4:1) produce instead recoiling holes with a kick velocity that is equal to the most likely value of the probability distribution function for randomly oriented spins (see Fig. 3 of \citealt{gue11}). In practice, this means that all recoiling holes will have speeds in excess of $100\,\kms$ and will be ejected from the system. This is a conservative assumption in the sense that it tends to minimize the actual number of relic IMBHs in the Galactic halo. We trace all $<$4:1 halo mergers following the VLII merger tree back to redshift 10. This was built by identifying all subhalos in each snapshot with the 6DFOF halo finder \citep{die06}, and linking them from one snapshot to the next by the id numbers of the shared particles. We identify all halos that would have been assigned a hole at infall, but had in their past a $<$4:1 major merger with another black hole-hosting system. As the kick velocity is always larger than the escape speed from the host, the hole is removed from the final catalog of Galactic IMBHs. 

\begin{figure*}
\centering
\includegraphics[width=0.48\textwidth]{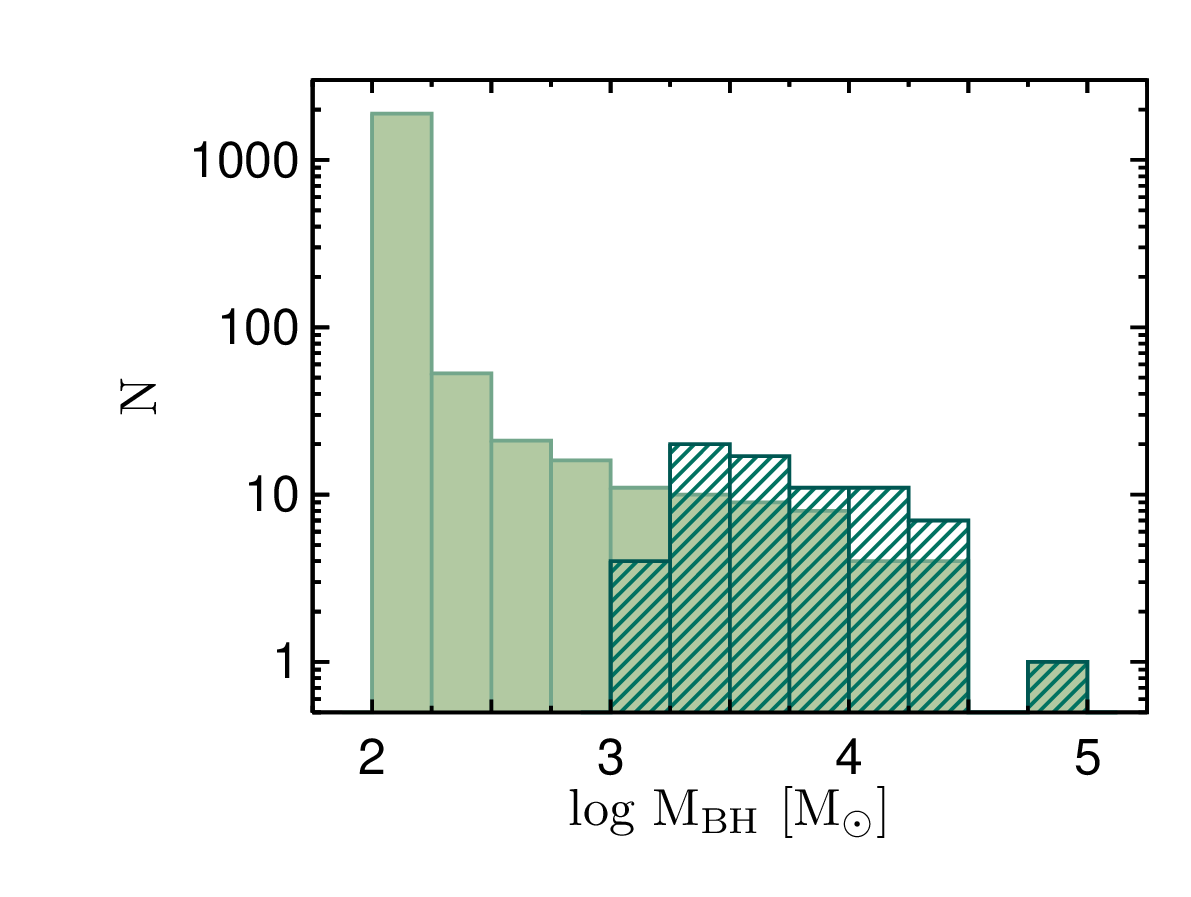}
\includegraphics[width=0.48\textwidth]{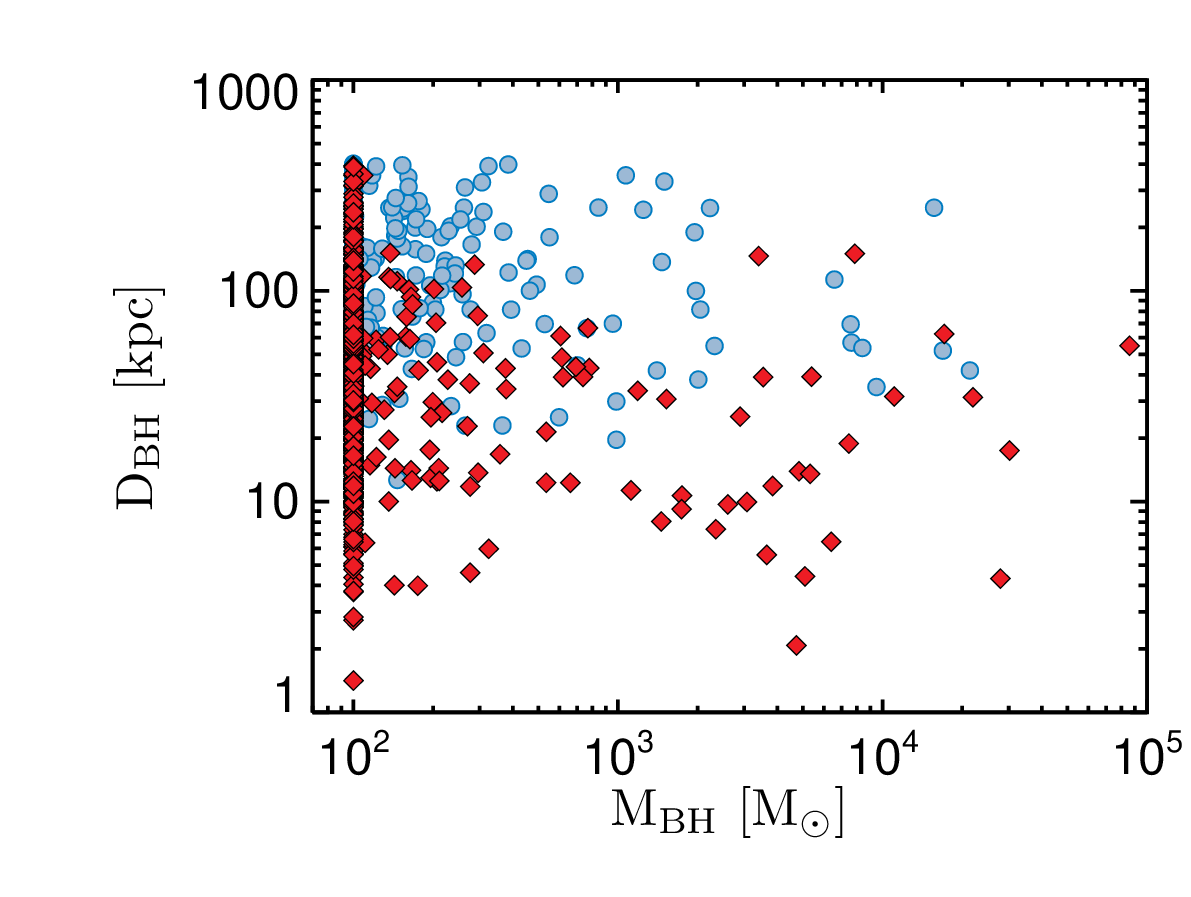}
\caption{{\it Left panel:} Mass function of relic IMBHs within $r_{200}$ in the two scenarios discussed in the text. 
{\it Solid histogram:} Pop III remnants. {\it Hatched histogram:} Direct collapse precursors. {\it Right panel:} Location of 
``naked'' ({\it red diamonds}) and ``clothed'' ({\it blue circles}) IMBHs in the $M_{\rm BH}$-Galactocentric distance plane for the Pop III 
model. Naked IMBHs are more concentrated towards the inner halo regions, a trend that is also observed in the direct collapse, massive
seed scenario.
}
\label{fig3}
\vspace{+0.cm}
\end{figure*}

\begin{table*}
\caption{\enspace Demography of Galactic IMBHs}\label{tab:multimodel}
\begin{tabular*}{\hsize}{@{\extracolsep{\fill}}lrcrrcrcr}
\\[-5pt]
\hline
\hline
\multicolumn{1}{l}{Model} & 
\multicolumn{1}{c}{$\sigma_m$} &  
\multicolumn{1}{c}{med$(M_{\rm halo})$} &
\multicolumn{1}{c}{$N_{\rm naked}$} &
\multicolumn{1}{c}{$N_{\rm clothed}$} &
\multicolumn{1}{c}{$N_{\rm massive}$} &
\multicolumn{1}{c}{min($M_{\rm BH}$)} &
\multicolumn{1}{c}{max($M_{\rm BH}$)} &
\multicolumn{1}{c}{$N_{\rm recoiled}$} \\
&
\multicolumn{1}{c}{[$\kms$]} &
\multicolumn{1}{c}{[$M_{\odot}$]} & & &
\multicolumn{1}{c}{($M_{\rm BH} > 5,000\,\msun$)} &
\multicolumn{1}{c}{[$M_{\odot}$]} &
\multicolumn{1}{c}{[$M_{\odot}$]} & \\
\hline
\\[-5pt]
Population III & $3$ & $5 \times 10^7$ & $798 $ & 
        $1234 $ & $21 $ & $100$ & $1.9 \times 10^5$ & $577$ \\ 
Direct Collapse & $12$ & $3 \times 10^9$ & $39 $ & 
        $33 $ & $37 $ & $1691$ & $1.9 \times 10^5$  & $5$ \\
\hline
\end{tabular*}
\tablecomments{Column 1 indicates the seeding scenario, columns 2, 3, 4, 5, 6, 7, 8, and 9 give the minimum subhalo line-of-sight stellar velocity dispersion required for hosting a central black hole, the median halo mass at infall, the number of ``naked" and ``clothed" Galactic IMBHs, the number of Galactic IMBHs more massive than $5000\,\msun$, the mass of the lightest and heaviest Galactic IMBHs, and the number of recoiled IMBHs, respectively.}
\vspace{+0.0cm}
\end{table*}

\begin{figure}
\centering
\includegraphics[width=0.48\textwidth]{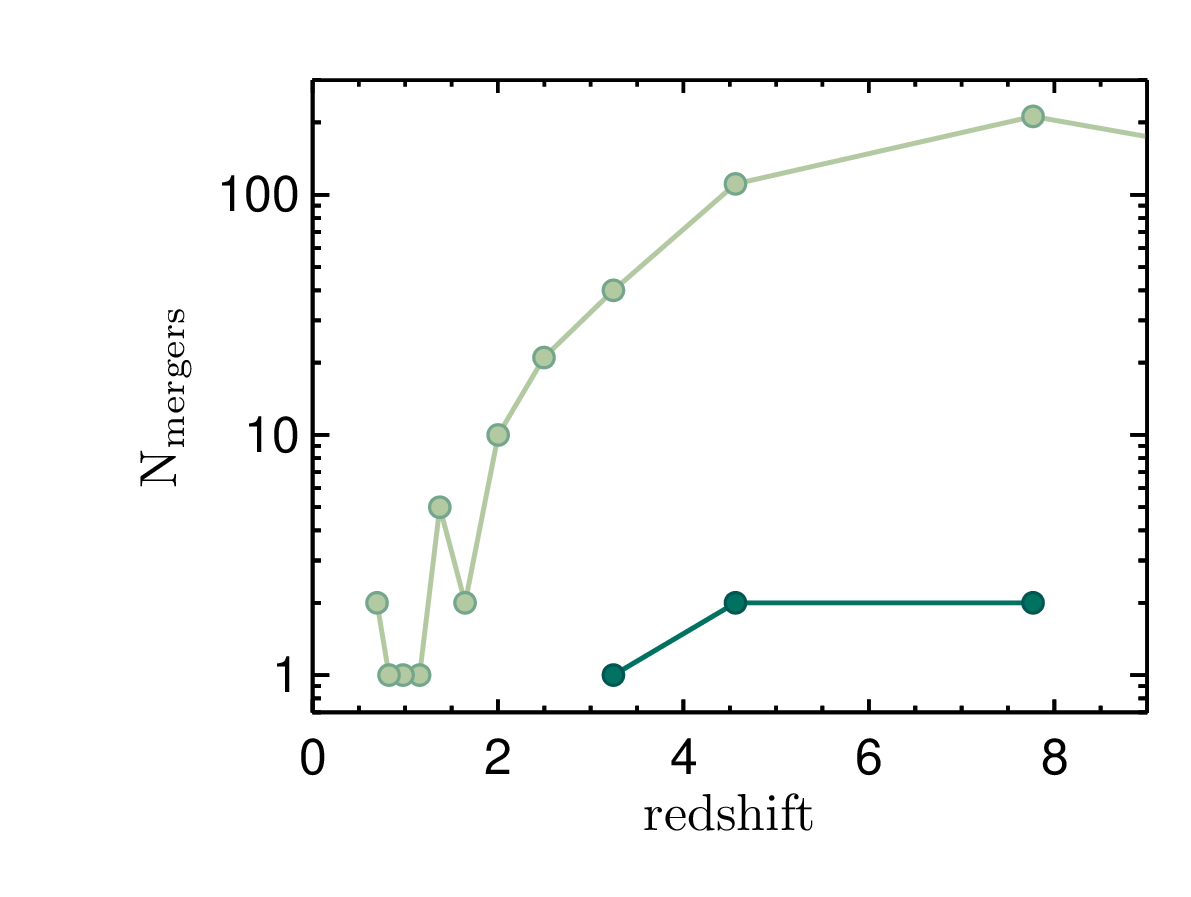}
\includegraphics[width=0.48\textwidth]{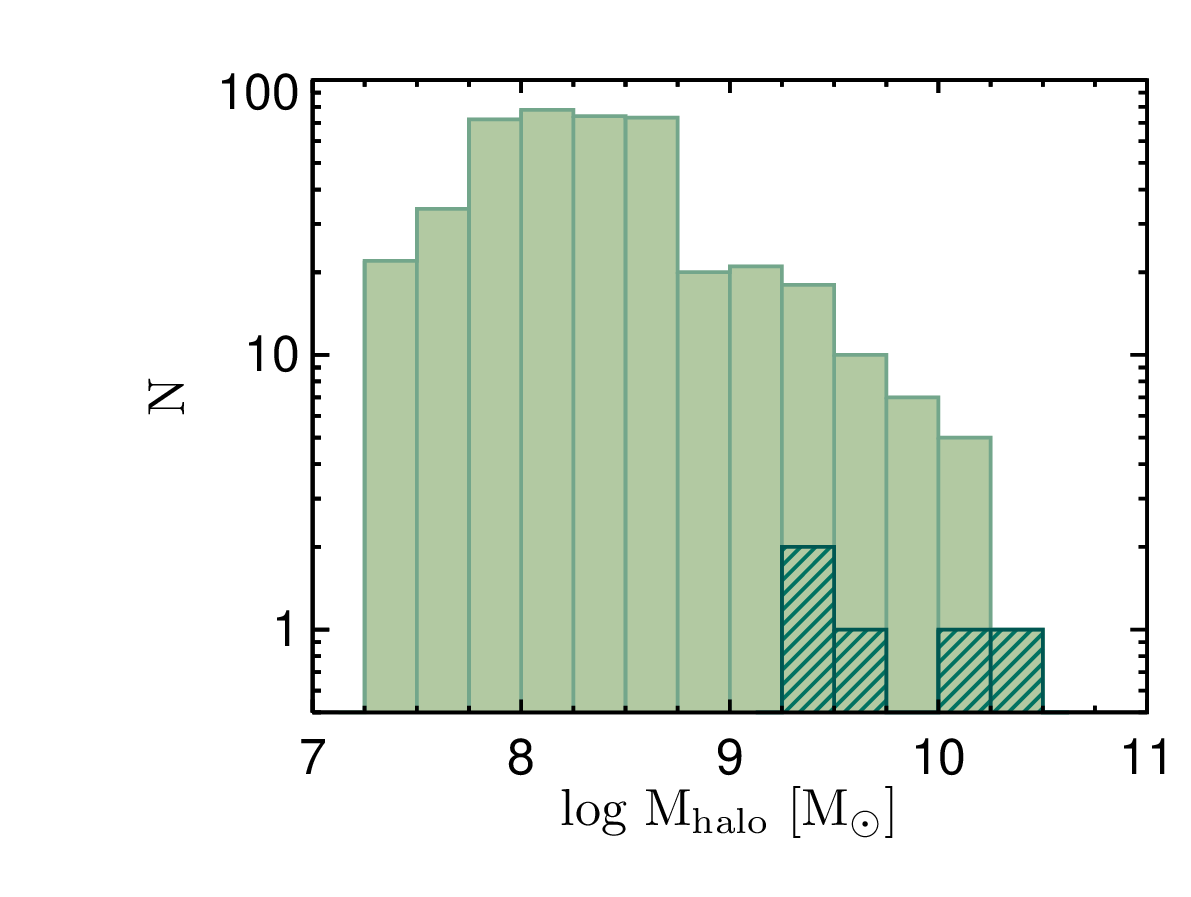}
\caption{{\it Top panel:} Number of subhalo major mergers that produce a recoiling IMBH, as a function of redshift. {\it Light green curve:} Pop III remnants model. 
{\it Dark green curve:} Direct collapse precursors. {\it Lower panel:} Mass distribution (at infall) of all the subhalos whose central black holes were ejected following a major merger. The color scheme is the same as in the top panel.
}
\label{fig4}
\vspace{+0.0cm}
\end{figure}

The size of the recoiling population is clearly very sensitive to the hole occupation fraction. Figure \ref{fig4} shows the total number of halo 
major mergers that produce recoiling IMBHs in the ``Pop III" and ``direct collapse" seed scenarios. The former produces 577 recoiling holes
(22\% of the total black hole population), mostly belonging to $\sim 10^8\,\msun$ subhalos at infall. The latter produces only 5 recoiling holes 
(6\% of the total) all from rather massive hosts. Note that it is because of recoils that the number of Galactic IMBHs heavier than a few thousand solar masses is actually larger in the direct collapse model (see Fig. \ref{fig3}). The demography of Galactic IMBHs in our two scenarios is summarized in Table \ref{tab:multimodel}.

\section{Spatial Distribution of IMBHs}

\begin{figure}
\centering
\includegraphics[width=0.48\textwidth]{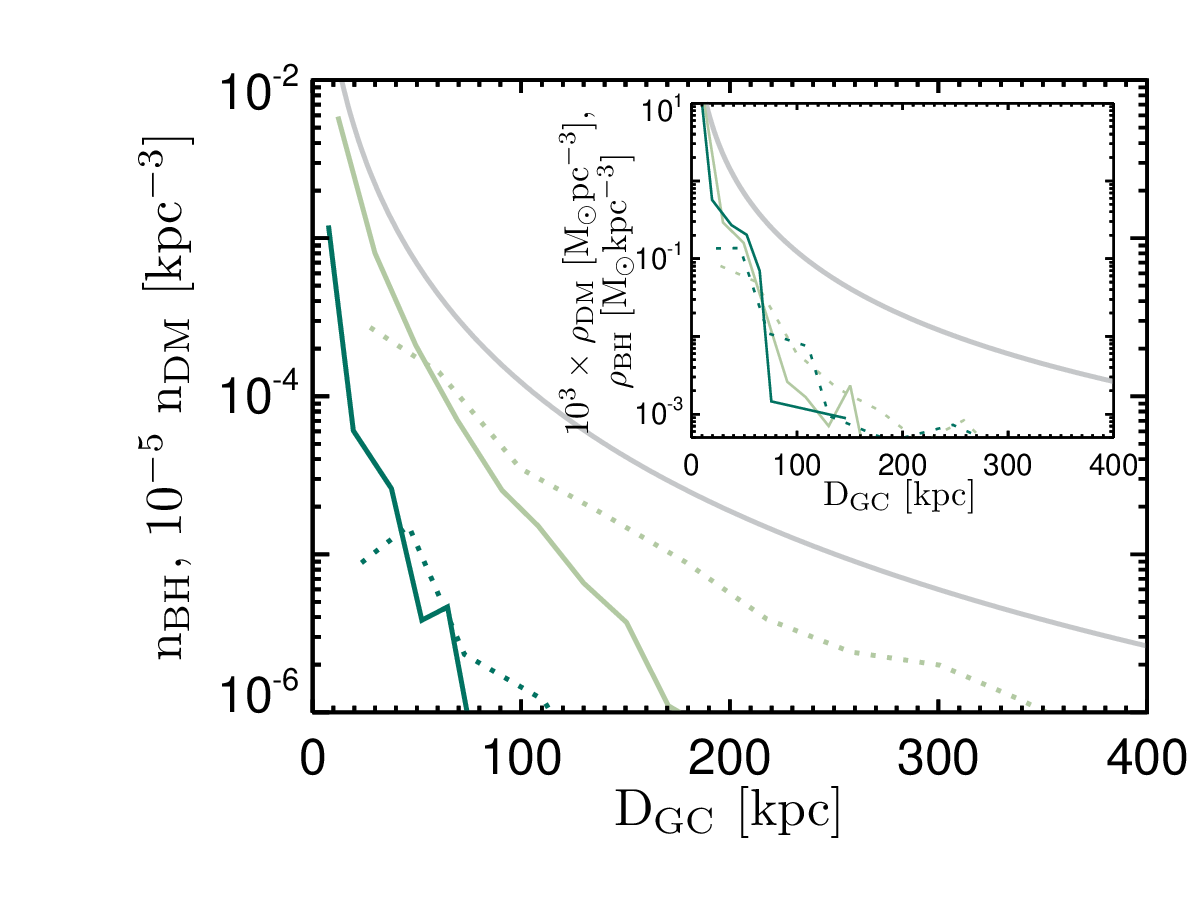}
\caption{Spatial distribution of Galactic IMBHs versus the dark matter density profile in VLII. {\it Main panel:} IMBH number density {\it Inset:} IMBH mass density. The mass density of the dark matter has been multiplied by a factor of 1,000 for comparison. {\it Light green:} Pop III remnants scenario, naked ({\it solid curves}) and clothed ({\it dotted curves}) IMBHs. {\it Dark green:} direct collapse progenitors scenario, naked ({\it solid curves}) and clothed ({\it dotted curves}) IMBHs.  
{\it Gray:} VLII dark matter density profile. The total number density of IMBHs follows the dark matter density profile, while their mass density does not. 
}
\label{fig5}
\vspace{+0.0cm}
\end{figure}

The spatial distribution of Galactic IMBHs today is shown in Figure \ref{fig5}. Naked holes dominate in the inner $\sim$ 50 kpc from the halo center. Their number density, however, plummets at larger distances, where most IMBHs are still hosted in surviving substructures. This is a  manifestation of the fact that subhalos orbiting closer to the dense central regions of the main host are more likely to be tidally disrupted, and leave their IMBHs exposed. In the outer halo regions, the lower background density makes substructures more resilient to tidal forces, allowing them to survive and retain their holes. A comparison of the IMBH number density to the density profile of the dark matter shows that the cumulative distribution of Galactic IMBHs (naked plus clothed) follows the dark matter profile rather well, but that each individual 
subpopulation does not. The inset in Figure \ref{fig5} shows that the IMBH mass density profile does not follow the density profile of the dark matter, as most of the mass in black holes is concentrated in the inner regions. 

Since naked IMBHs more massive than dark matter particles do not experience proper dynamical friction in our simulation, we have estimated 
a posteriori the effect of black hole orbital decay in two different ways -- using the slowly-decaying circular orbit approximation, as well as computing the instantaneous deceleration from dynamical friction \citep{bin08} in each available VLII snapshot. We find that, even for the heaviest IMBHs in the highest density regions of our simulation, dynamical friction timescales exceed the Hubble time at all points along the
orbit. A similar conclusion was reached by \citet{ole09} in their study. Of course, our N-body simulation does not account for possible hydrodynamical effects and interactions with the galaxy stellar disk in the very center of the VLII halo.

\section{Discussion}

We have used a particle tagging technique to populate the subgalactic building blocks of a present-day MW-sized galaxy with black holes, and assess the size, properties, and detectability of the leftover population.  Our approach combines a computationally expensive, 
extreme-resolution, N-body simulation like VLII, in which structures grow ab initio in a fully cosmological framework, with simple prescriptions for seeding, in post-processing, progenitor subhalos with IMBHs. Insofar baryonic material does not appreciably perturb the collisionless dynamics of the N-body component, the dynamical association of black holes with individual particles in the N-body component should correctly reproduce the spatial and kinematic properties of IMBHs in galaxy halos, in particular of the naked subpopulation whose host satellite galaxies have been totally destroyed after infall. This level of detail is unavailable to a standard merger-tree approach. 

We have discussed two simple models of seed hole formation that may be illustrative of more realistic growth scenarios, a ``Pop III remnant" model in which small-mass black holes populate subhalos with stellar velocity dispersion as low as $3\,\kms$, and a ``direct collapse precursor" model in which holes become very rare in systems with stellar velocity dispersion below $12\,\kms$, and are more massive. It is important to keep in mind that the Pop III route does not necessarily require the formation of extremely massive stars from gas of primordial composition. Already at metallicities $Z\lta 0.01\,Z_\odot$, the mass loss rates of massive stars through radiatively driven stellar winds are predicted to be rather small \citep[with rates decreasing with metallicity as $\dot M\propto Z^{0.69}$, see][]{vin01}. If low-metallicity stars above $50\,\msun$ collapse to black holes after losing only a small fraction of their initial mass, then IMBHs with masses above the $5-20\,\msun$ range of known ``stellar-mass" holes may be the inevitable end product of early star formation. A standard Salpeter initial mass function (IMF), $dN/dm_*\propto m_*^{-2.35}$ in the range $1-100\,\msun$, would produce an IMBH every $1000\,\msun$ of stars formed. A top-heavy IMF like that found in recent simulations of primordial star formation by \citet{sta12}, $dN/dm_*\propto m_*^{-0.17}$, over the same $1-100\,\msun$ mass range, would generate an IMBH every $100\,\msun$ of stars formed. Note that $1000\,\msun$ of stars in a $5\times 10^7\,\msun$ subhalo imply a star formation efficiency that is much too high to match the observed dearth of faint Milky Way satellites (e.g. Rashkov \etal 2012), {\it and a top-heavy IMF must then be invoked for consistency.}

\begin{figure}
\centering
\includegraphics[width=0.48\textwidth]{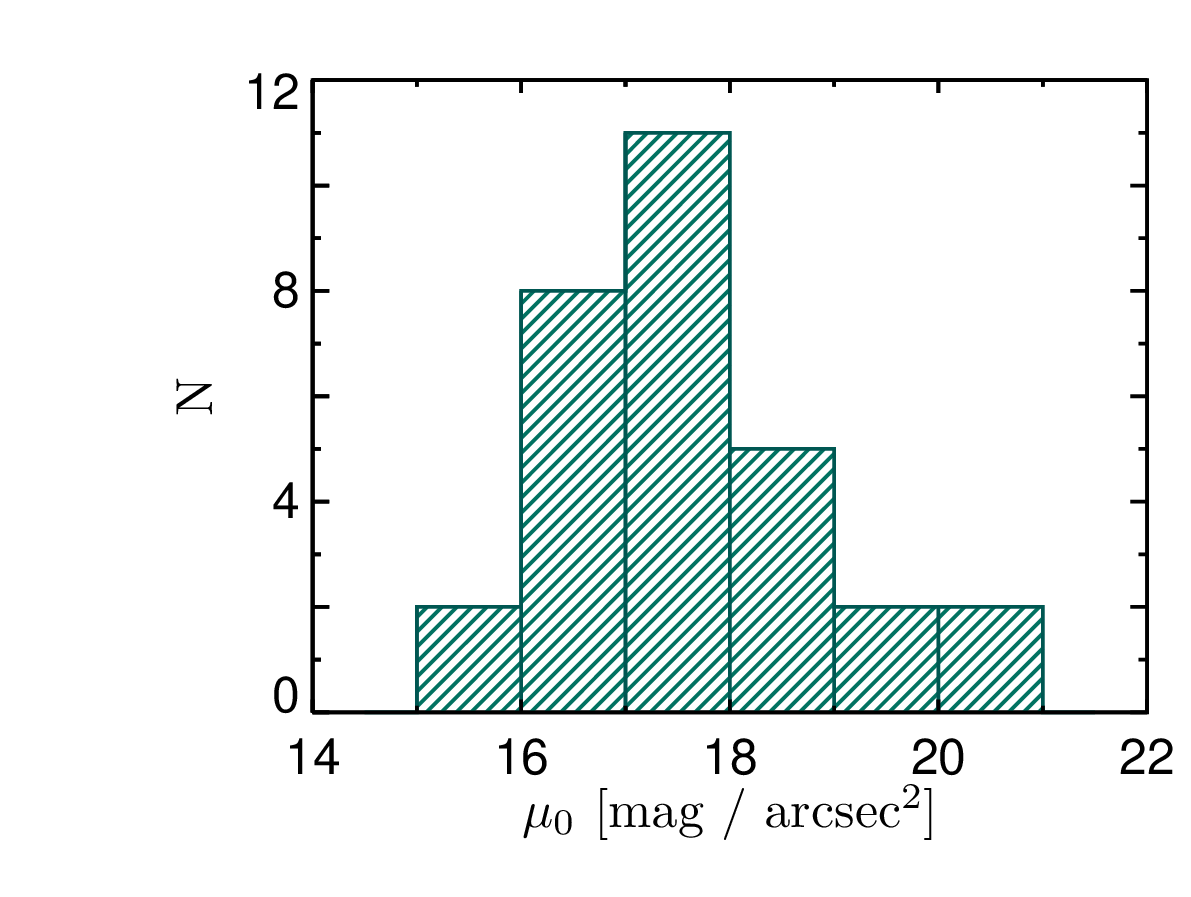}
\caption{Distribution of V-band peak surface brightness (within the central 1 arcsec) of stellar cusps around naked IMBHs in the ``direct collapse" model. Only 20$\%$ of the initial stellar mass (equal to $2\times M_{\rm BH}$) is assumed to be retained today. To account for the partial SDSS sky coverage, the expected numbers should be scaled down by a factor of $\sim$5.
}
\label{fig6}
\vspace{+0.0cm}
\end{figure}
 
Galactic IMBHs may light up if they pass through dense regions of the galaxy and accrete from the interstellar medium \citep{vol05}. IMBHs present in the lensing galaxy of multiply-imaged background QSOs will produce monopole- or dipole-like distortions in the surface brightness of the QSO images and may be detectable by next generation submillimeter telescopes with high angular resolution \citep{ino13}.  It is interesting at this stage to discuss the observability of the stellar cusps that may accompany naked IMBHs wandering in the Milky Way halo. After the complete disruption of their host subhalo, naked IMBHs are still expected to carry with them a Bahcall-Wolf stellar cusp \citep{mer09} which, upon formation and in the absence of a kick from gravitational recoil, has a mass of order the mass of the hole and an extent of order the hole's radius of influence,
\begin{equation}
r_{\rm BH} = GM_{\rm BH}/\sigma_*^2.
\end{equation}
Throughout their dynamical evolution, these stellar clusters will lose mass by relaxation and expansion, as well as tidal disruptions by the 
black hole \citep{kom08,ole12}. The relaxation timescale, 
\begin{equation}
t_{\rm r} \approx 10^9 \left(\frac{M_{\rm BH}}{10^5 M_{\odot}}\right)^{5/4} \rm{yr},
\end{equation}
ranges from several Myr for the smallest IMBHs to a Gyr for the largest. As IMBHs get accreted by the Milky Way early in the history of the Universe, the above processes can cause a 60-80\% fractional loss of stellar mass by the present day \citep{ole12}. At the same time, the physical extent of the clusters is expected to grow like $t^{1/3}$ after the first relaxation timescale \citep{ole12}, resulting in a total expansion by a factor of 16 (for $M_{\rm BH} \sim 10^3 M_{\odot}$) to 3 (for 
$M_{\rm BH} \sim 5 \times 10^4 M_{\odot}$) over a Hubble time. Assuming that the initial extent of the clusters is dictated by the IMBH's sphere of influence at infall, their physical extent today would range between 0.5 and 1.0 pc. Despite their expansion, stellar clusters surrounding IMBHs would still be very compact today and may appear point-like or extended depending on their distance. Photometrically, they would have colors similar to those of old stars, while spectroscopically they would be distinguishable by the high velocity dispersions.

\begin{figure}
\vspace{+0.cm}
\centering
\includegraphics[width=0.48\textwidth]{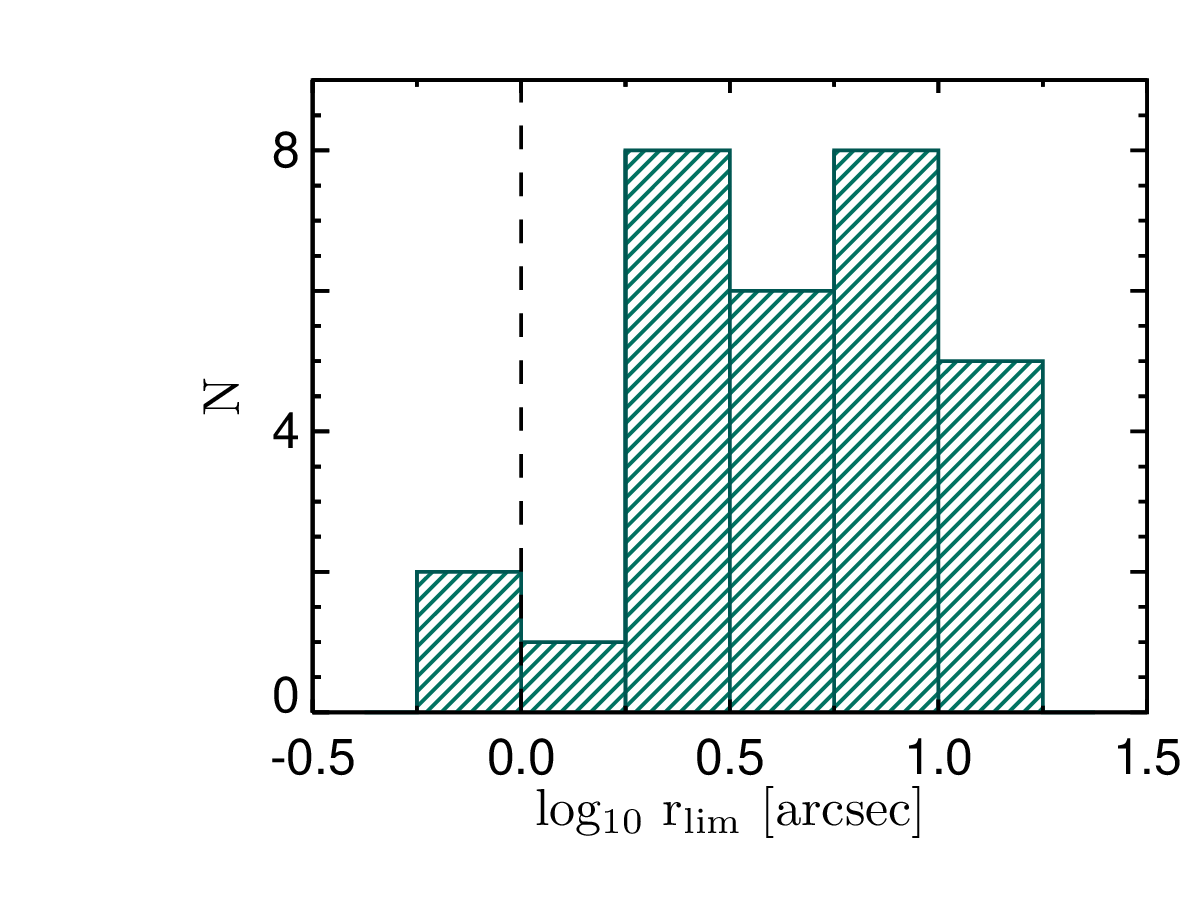}
\caption{Angular size of stellar clusters around Galactic ``naked" IMBHs in the ``direct collapse" model (see text for details).
The angular extent is calculated for an observer at $d_{\rm GC}=8$ kpc from the center of the VLII halo.
{\it Dashed line:} angular resolution limit of current ground-based surveys.
}
\label{fig7}
\vspace{+0.cm}
\end{figure}

We have modelled the apparent brightness distribution of stellar cusps around naked IMBHs in the direct collapse scenario, which produces the largest number of massive IMBHs today. We assume an initial stellar mass that is twice the mass of the black hole and a present-day stellar mass-to-light ratio of 4 \citep[as appropriate for an old stellar population with a Salpeter IMF - see][for details]{ras12}. The mass density in the star clusters is expected to follow a steep power law with drop-off power $\alpha \sim$ -2.15. Given these physical characteristics, we calculate the peak surface brightness (within a typical SDSS seeing element of angular size $\sim$ 1 arcsec) of the clusters as would be seen by a mock observer located 8 kpc from the center of the VLII main halo. The distribution is shown in Figure \ref{fig6}. The angular size out to which the brightness profile will be brighter than 25 mag arcsec$^{-2}$ is shown in Figure \ref{fig7}. Most clusters would be resolvable, as their angular extent spans 2 to 10 arcseconds.
In these calculations, we have accounted for the fact that the SDSS points away from the Galactic Center by excluding IMBHs within 8 kpc of the VLII halo center. The resulting distribution of peak brightness for the IMBH stellar clusters would have to be further scaled down by a factor of 5, to account for the limited sky coverage of the SDSS. After accounting for these corrections, only a handful of clusters should be detectable with current surveys. This is consistent with a recent search for stellar clusters surrounding recoiled IMBHs in the SDSS source catalog that resulted in an upper limit of 100 candidates in the Milky Way halo \citep{ole12}. We have also estimated the proper motion of IMBHs on the sky. Having tangential velocities ranging from 0 to 250 $\kms$ as seen from an observer at 8 kpc from the VLII halo center, and distances of up to 100 kpc, their proper motions fall in the range between 0.1 and 10 milliarcsec year$^{-1}$. This is similar to measured proper motions of distant halo stars and may be currently detectable with the {\it Hubble Space Telescope} \citep{soh12}, and future missions such as {\it Gaia}.

\begin{figure}
\vspace{+0.cm}
\centering
\includegraphics[width=0.48\textwidth]{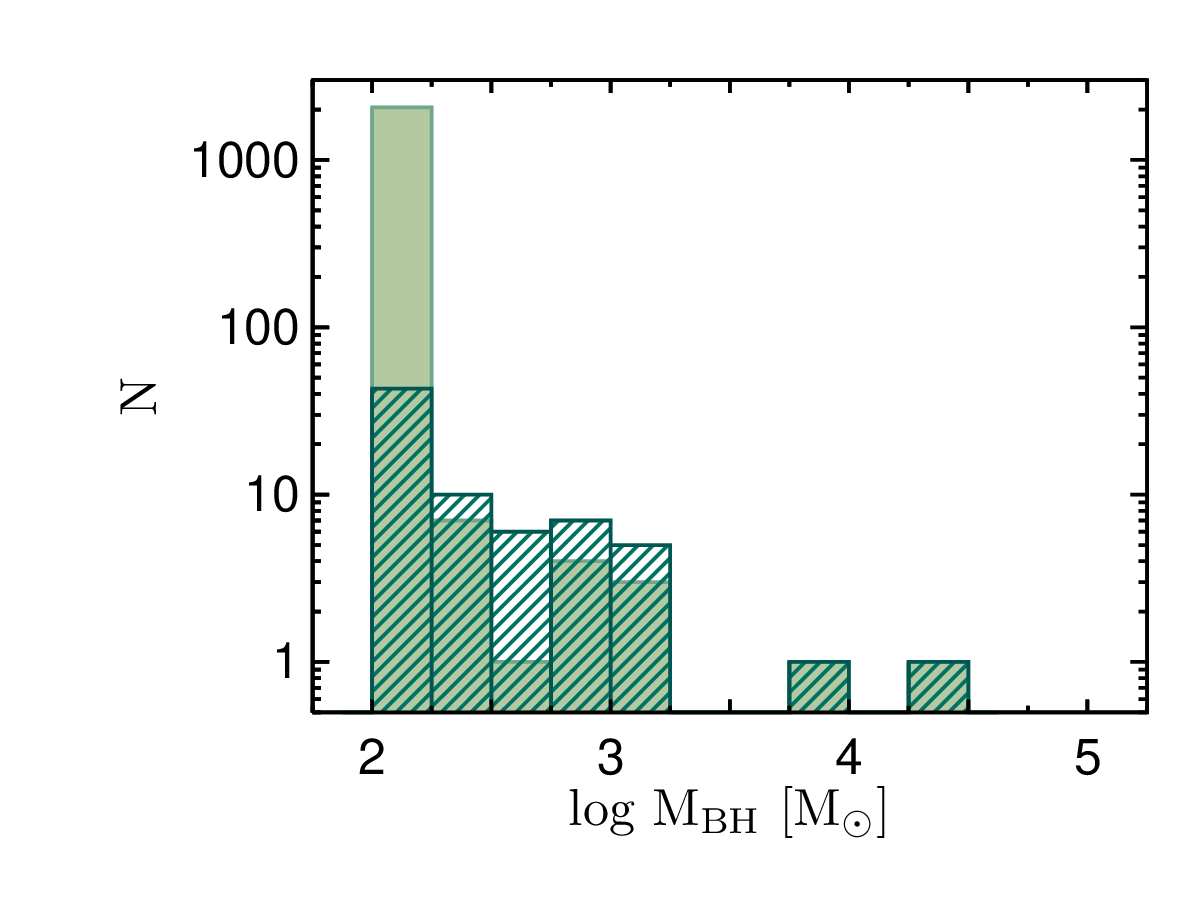}
\caption{
{The impact of the steeper $M_{\rm BH}-\sigma_*$ relation from \citet{mcc13} on the mass function of relic IMBHs. {\it Solid histogram:} Pop III remnants. {\it Hatched histogram:} Direct collapse 
precursors. Comparison with Figure 3 clearly shows the significant reduction in the predicted masses of IMBHs.} 
}
\label{fig8}
\vspace{+0.cm}
\end{figure}

Finally, we assess the impact of a steeper $M_{\rm BH}-\sigma_*$ relation as measured by \citet{mcc13} in Figure \ref{fig8}. Like before, we tag all IMBHs 
according such relation and then remove from the final catalog all recoiling holes. The steeper slope (with exponent equal to 5.64 instead of 4) produces  much 
lighter black holes, by about a factor of 30, over the whole spectrum of masses. This effect is highlighted by the migration of most holes in the Pop III scenario 
to the lightest (leftmost) bin in the histogram in Figure \ref{fig8}. This 30$\times$ reduction in mass of the largest IMBHs results in a 30$\times$ reduction in 
the initial size of their stellar cusps, and in a large reduction of the stellar relaxation timescale, making them too compact and faint to be detectable with 
current surveys. A similar result would hold for those models in which the IMBHs hosted by dwarf satellites fall below the $M_{\rm BH}-\sigma_*$ relation, as 
in \citet{van10}.

\acknowledgments
We acknowledge useful discussions on the topics of this paper with Z. Haiman, S. Tremaine, and M. Volonteri.
Support for this work was provided by the NSF through grant OIA-1124453, and by NASA through grant NNX12AF87G.

\end{document}